\documentstyle[aps,prl,floats,epsfig,twocolumn]{revtex}

\input{tcilatex}

\begin{document}

{\bf Comment on ``Low Temperature Magnetic Instabilities in Triply Charged
Fulleride Polymers''}

Recently, Ar\v{c}on {\it et al}.\cite{denis} reported ESR\ studies of the
polymer phase (PP) of Na$_{2}$Rb$_{0.3}$Cs$_{0.7}$C$_{60}$ fulleride. It was
claimed that this phase is a quasi-one-dimensional metal above 45 K with a
spin-gap below this temperature\ and has antiferromagnetic(AF) order below
15 K,\ that is evidenced by antiferromagnetic resonance(AFMR).

For the understanding of the rich physics of fullerides it is important to
identify the different ground states. ESR has proven to be a useful
technique for this purpose. However, since it is a very sensitive probe, it
can detect a multitude of spin species and it is not straightforward to
identify their origin, especially in a system like Na$_{2}$Rb$_{x}$Cs$_{1-x}$%
C$_{60}$\ with three dopants, when one part of the sample polymerizes but
the majority does not. The observation of a low dimensional instability in
the single bonded PP would be a novel and important result. Nevertheless, in
this Comment we argue that Na$_{2}$Rb$_{0.3}$Cs$_{0.7}$C$_{60}$ is not a
good choice for this purpose since, as we show, the samples used in Ref.\cite
{denis} are inhomogeneous. We point out that recent results on the PP of Na$%
_{2}$CsC$_{60}$ contradicts the observation of low dimensional instabilities
in Na$_{2}$Rb$_{0.3}$Cs$_{0.7}$C$_{60}$.

The ESR\ signal of Ref.\cite{denis} at 285 K\ consists of two components as
read from Fig. 1 of Ref.\cite{denis}. For the narrower component the
peak-to-peak linewidth $\Delta H_{pp}\approx 60$ G . This is not the signal
of the simple cubic({\it sc}) phase of Na$_{2}$Rb$_{0.3}$Cs$_{0.7}$C$_{60}$
as $\Delta H_{pp}\approx 250$ G is expected for this phase from measurements
on the {\it sc} phases of Na$_{2}$CsC$_{60}$ ($\Delta H_{pp}\approx 370$ G 
\cite{na2csprl}) and Na$_{2}$RbC$_{60}$ ($\Delta H_{pp}\approx 40$ G\cite
{denisprb}\cite{kirch99}) at 285 K.{\em \ }A broader component in the ESR
signal in Fig. 1 of Ref.\cite{denis} is indeed visible, as a tail around 0.3
T and is better seen in the integrated spectrum at 220 K with $w\approx $ $%
200$ G ($\Delta H_{pp}\approx 230$ G). This multi-component nature of the
ESR signal suggests a phase separation rather than local disorder as the
latter would be averaged by conduction electrons. The narrow(broad) ESR\
component may come from Rb(Cs) rich and Cs(Rb) poor parts of the sample. The
non-stoichiometry of the compound may be the reason for this phase
separation that affects the PP, as well. It is not documented in Ref.\cite
{denis} whether contrast between Rb and Cs in the X-ray experiment allows
the exclusion of the above suggested phase separation.

In addition to the apparent phase separation of the sample used in the
experiment, we could not reproduce the temperature dependence of the ESR
intensity in Fig. 3a and c from the raw ESR spectra of Fig. 2 of Ref.\cite
{denis} and from the w$_{T}$\ values of Fig. 3b and d. It appears as if the
broadening of the ESR line below 45 K (that reduces the amplitude of the ESR
signal of the powder distribution) was not taken into account of the
calculation of the ESR\ intensity. This leads us to question the observation
of a spin-gap below 45 K.

Below 15 K, Ar\v{c}on {\it et al}.\cite{denis}\ attributes the AFMR to an
emerging ESR signal in high frequency (HF)-ESR which is absent in X-band.
The HF-ESR linewidth and field shift from the ESR signal of the PP allowed
the calculation of a reasonable value of spin-flop(SF) field. This is
insufficient for the unambiguous identification of an AFMR, that requires
ESR measurements at least at two high frequencies above the SF field\cite
{afmr} and the observation of a decreasing linewidth and field shift with
increasing ESR frequency. It would be important to see the result of these
experiments that could be performed at NHMFL\cite{tallahassee}.

In our opinion, it is very intriguing that the presence of superconducting
phase in the sample is neglected in the discussion of Ref.\cite{denis}. 81
\% of Na$_{2}$Rb$_{0.3}$Cs$_{0.7}$C$_{60}$ is in the {\it sc} phase and is a
superconductor with T$_{c}\approx $ 10 K\cite{yildirimssc95}. The so called
vortex noise due to this significant amount of superconducting phase
probably prevents reliable conclusions from X-band data (not shown in Ref. 
\cite{denis}) below T$_{c}$. Thus it can not be decided whether the signal
that is observed in the HF experiment is present or not in X-band. Moreover,
it can not be excluded that the signal observed below 15 K in the HF
experiment and attributed to the AFM phase is the ESR signal of the residual 
{\it sc} phase. We estimate that $w\approx 7${\em \ }mT for the {\it sc}
phase at 15 K from results on Na$_{2}$CsC$_{60}$\cite{na2csprl} and Na$_{2}$%
RbC$_{60}$\cite{kirch99}. This ESR line is expected to narrow below T$_{c}$
similarly to the situation encountered in K$_{3}$C$_{60}$\cite{norbi}, which
may lead to a linewidth of the {\it sc} phase similar to that assigned to
the AFMR\ signal.

In summary, we have shown that the analysis and discussion of the
experimental data in Ref.\cite{denis} is ambiguous and full reconsideration
of the results is necessary. The main reason of the authors of Ref.\cite
{denis} to expect electronic instabilities in the PP of Na$_{2}$Rb$_{0.3}$Cs$%
_{0.7}$C$_{60}$ is the expanded interchain distance in comparison with e.g.
the PP\ of Na$_{2}$RbC$_{60}$, the latter being a metal till 4 K. It has
recently been shown that the even further expanded lattice size Na$_{2}$CsC$%
_{60}$ polymer\cite{latticecomment}\cite{margadonnaJSSC} is a metal until
low temperatures\cite{na2csprl}. Thus, the ground state of the single bonded
fulleride polymers is more likely a metal than a spin density wave insulator.

Ferenc Simon, Slaven Garaj and L\'{a}szl\'{o} Forr\'{o}

\'{E}cole Polytechnique F\'{e}d\'{e}rale de Lausanne CH-1015, Lausanne,
Switzerland


\begin{references}
\bibitem{latticecomment}  The lattice constants of Na$_{2}$Rb$_{0.3}$Cs$%
_{0.7}$C$_{60}$ were not presented in Ref.\cite{denis}.
\end{references}
\end{document}